\documentstyle[seceq,preprint,epsf]{ptptex}
\newcommand{\beq}{\begin{equation}}
\newcommand{\eeq}{\end{equation}}
\newcommand{\kk}{$\mathbf k$}
\newcommand{\apot}{$\mathbf A$}
\preprintnumber[3cm]{HLRZ~05/98\\ WUP-TH~05/98\\ HUB-EP~98/11}
\title{Probing the QCD Vacuum  with Static Sources\\
 in Maximal Abelian Projection } 
\author{Gunnar~S. {\sc Bali}$^c$\footnote{E-mail address: bali@pha1.physik.hu-berlin.de}, 
Christoph {\sc Schlichter}$^{a}$, 
and Klaus {\sc Schilling}$^{a,b}$\footnote{Talk presented at YKIS97, E-Mail address:
schillin@theorie.physik.uni-wuppertal.de}
}

\inst{
$^a$Fachbereich Physik, Bergische
        Universit\"at, D-42097 Wuppertal, Germany\\
$^b$HLRZ c/o Forschungszentrum J\"ulich,
        D-52425 J\"ulich and DESY,  Germany\\
$^c$Institut f\"ur  Physik, Humboldt Universit\"at, Invalidenstr.~110,\\
D-10115 Berlin, Germany
}

\recdate{ \today } 

\abst{ Various field strength correlators are investigated in the
maximal Abelian projection of pure SU(2) lattice gauge theory. High
precision measurements of the colour fields, monopole currents, their
curls and divergences allow for detailed checks of the dual
superconductor scenario. On this basis, we perform a Ginzburg-Landau
type analysis of the flux tube profile from which we derive the size
of the penetration length, $\lambda = 0.16(2)$~fm, and coherence length
of the monopole condensate wave function, $\xi = 0.27(3)$~fm. The ratio of
these numbers is $\kappa = \lambda/\xi = 0.59(13)$ which is below the value
$1/\sqrt{2}$ where type II superconductivity sets in.}

\begin{document}
\maketitle
\section{Introduction}
 The mechanism of colour confinement is deeply connected with the
structure of the QCD vacuum. In the scenario of {\sc 't Hooft} and {\sc
Mandelstam}, confinement is viewed as the strong interaction analogue
of the well known Mei\ss ner effect, in a dual
superconductor\cite{thooft,mandelstam}.

As we heard in the introductory lectures by Prof.\ DiGiacomo during
this workshop, the underlying idea is that the QCD vacuum is filled by
a chromo-magnetic monopole condensate.  Thus, if we insert static
colour charges in form of a heavy quark-antiquark pair, $Q\overline{Q}$,
the monopole condensate will expel the chromo-electric field from the
vacuum.  This would then provide the mechanism for chromo-electric
flux tube formation and the ensuing  linearly rising potential.

On the level of the potential this picture has been verified with
progressively refined techniques of lattice gauge theory, ever since
the seminal paper of {\sc Creutz}\cite{creutz}, both in pure gauge
theory\cite{potential} and full QCD\cite{potential_qcd}.

We would expect that a detailed determination of the flux tube profile
between static quarks offers additional insight into the understanding
of quark confinement. According to the above scenario, we should observe a
 normal conducting vortex string that distorts the surrounding
superconducting vacuum inside a penetration region of size
characterised by the inverse ``dual photon'' mass,
$\lambda $.  It is this very transition region between the normal
conducting vortex and the undistorted vacuum which is supposed to
reveal the interesting physics.

  In particular one would hope to gain, from a study of such surface
phenomena at the boundary of a dual superconductor, another 
insight on the second physical scale of spontaneous symmetry breaking,
namely the coherence length $\xi$ which is related to the Abelian
Higgs mass behind the monopole condensate.  This prospect would be
highly welcome, as it appears not at all straightforward to attain the
Higgs mass directly from correlators between monopole
creation/annihilation operators {\it in the purely gluonic
sector}\footnote{See the lectures of A. DiGiacomo in this volume.}.
Needless to say, though, that the viability of the
response-to-source-insertion approach needs to be demonstrated.

It is to be noted that a numerical investigation of the {\sc 't
Hooft-Mandelstam} scenario requires recourse to an Abelian gauge, such
as the maximal Abelian gauge projection (MAGP) proposed in
Ref.\cite{mag}, in order to make contact to Abelian mo\-no\-poles.  In an
impressive 
series of studies the Kanazawa group has established that MAGP indeed
accounts for most of the string tension\cite{suzukimag}. Moreover, it
was shown recently that uncertainties on this result, due to gauge
ambiguities in the actual projection procedure on the lattice, can be
largely excluded; in fact it was shown that the Abelian part of the
string tension accounts for 92 \% of the confinement part in the
static lattice potential\cite{mmp}.

\begin{figure}
\vskip .5truecm
\epsfxsize=12.5truecm
\centerline{\epsfbox{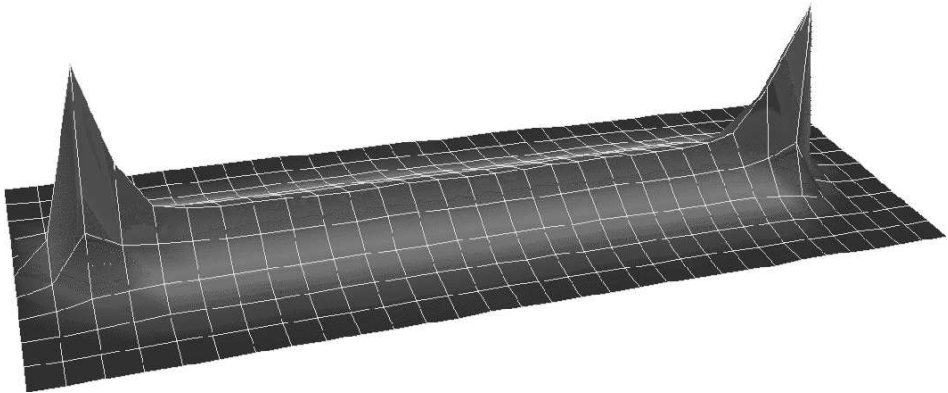}}
\vskip 1truecm
\caption{Energy distribution within a flux tube of length
$2$~fm between a $Q\overline{Q}$ pair
in MAGP.} 
\label{fluxg}
\end{figure}

Another important observation is that expectation values of field
distributions suffer much less from stochastic fluctuations after
MAGP\cite{diss}.  As we can demonstrate in Fig.~\ref{fluxg}, this
feature enables us to study string formation
over separations as large as $2$ fm in SU(2) gauge
theory. This is to be compared to standard SU(2) analyses, where so
far energy density distributions of the flux tube disappeared in the
noise at string elongations beyond $0.7$ fm\cite{flux}.

First attempts to determine the transverse flux tube profiles in MAGP
were launched some years ago in a number  of pioneering papers by {\sc
Haymaker}\cite{haymaker} et al.  as well as by the Bari\cite{bari} and
Kanazawa\cite{kanazawa} groups.  However, these authors had to work
with relatively small lattices and $Q\overline{Q}$ separations where the
flux tube has not yet fully developed, or non-optimised
noise-reduction techniques; as a result previous estimates for the
penetration length $\lambda$ and the Ginzburg-Landau coherence length
$\xi $\cite{GL} suffered from uncontrolled  systematic errors
and were by far not conclusive.

In the work reported here which is based on the thesis of one of us
(Ch.S.), we will try to go much beyond these early attempts, by
performing a simulation on a $32^4$ lattice at $\beta = 2.5115$ (with
the SU(2) Wilson action). The salient features of our analysis are:
(a) we are well positioned in the scaling regime; (b) we operate at
fairly fine lattice spacing, $a = 0.086$ fm as to resolve the
penetration region (estimated from the string tension $\sqrt{\kappa} =
440$ MeV); (c) we are working on a lattice large enough to attain $1$
fm separations between the sources.

We thus expect to meet a fair chance for establishing an observational
window towards  the boundary phenomena in quest, being safely
located between the
Scylla and Charybdis regimes of lattice artefacts and poor
signal-to-noise ratios.

\section{Measuring the Flux Tube}
Our present work is based on 116 independent SU(2) gauge
configurations that are gauge fixed and projected onto maximal Abelian
gauge according to the procedures described in Ref.\cite{mmp}.

After MAGP, the static $Q\overline{Q}$ sources are implemented by
 considering smeared $R \times T$ Wilson loops, $W(R,T)$, placed
 between time slices $0$ and $T$ on the lattice.  The field
 distributions around these static objects are determined by measuring
 correlations with suitable operators, say probes $O({\mathbf x},
 T/2)$. Thus, by varying $\mathbf x$ we can survey the environment of
 two static sources separated by a physical distance $Ra$ from each
 other.  Depending on the choice of the local probe $O$ we can measure
 the distributions of ${\mathbf E}$, ${\mathbf B}$, magnetic current
 ${\mathbf k}$ and their {\it curls, divergences} etc.

The physics information is retrieved from correlators of type,
\begin{equation}\label{fluxcorrelator}
\langle O ({\mathbf x},T) \rangle = \frac{\langle O ({\mathbf
x},T/2)W(R,T)\rangle}{\langle W(R,T) \rangle}-\langle O\rangle,
\end{equation}
in the limit of large $T$.
Note that we use smeared Wilson loops in order   to 
achieve  good signals (plateaus)
at {\it small} values of
$T$\cite{flux,new}. 

\subsection{Prerequisites: Checking Dual London Equations}
\begin{figure}[h]
\vskip - 2truecm
\epsfxsize=14truecm
\centerline{\epsfbox{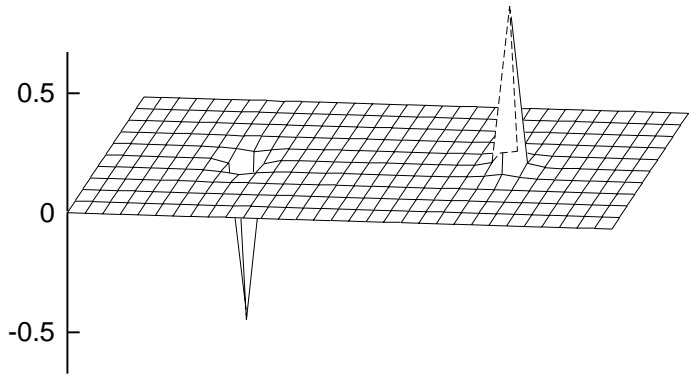}}
\vskip - 1.5truecm
\caption{Checking the localisations of $\mbox{div}\,\mathbf E $  on
source and sink, at $R = 15$.}\label{div}
\end{figure}

 At the outset we might wonder whether we are sensitive enough to spot
the origin of the chromo-electric fields to the locations of the
$Q\overline{Q}$ pair. In fact we find $\mbox{div}\, {\mathbf E}$ to vanish
nicely outside these sources, as illustrated in Fig.~\ref{div}.

An important ingredient of the dual theory is given by the magnetic
monopole current, \kk , which we assume to be defined on our lattice by
the standard {\sc DeGrand-Toussaint} prescription\cite{degrand}.
The expectation is, of course, that this current is a
solenoidal (i.e.\ azimuthal) supercurrent, stabilising the
normal conducting vortex core, and fulfilling the dual Amp\`ere
Law\footnote{Our $\mathbf k$
is $2\pi$ times the expression of Ref.\cite{degrand}.},
\begin{equation}
 {\mathbf k} = \mbox{curl}\, {\mathbf E}
     \label{dualamp}
\end{equation}

\begin{figure}[t]
\epsfxsize=12.5truecm
\centerline{\epsfbox{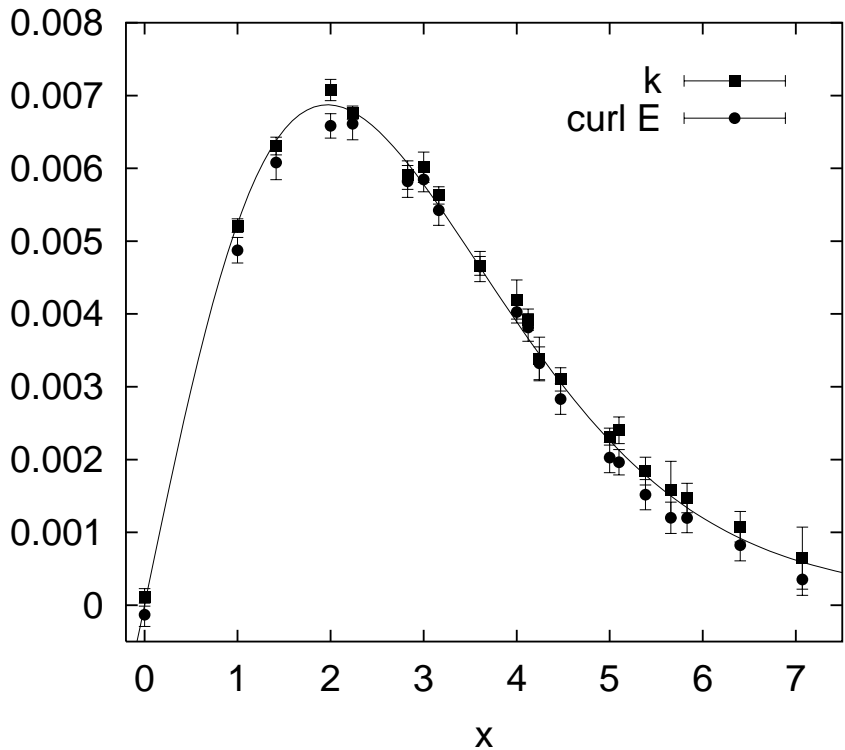}}
\caption{Checking the dual Amp\`ere Law, Eq.~(\protect\ref{dualamp}).}\label{ampere}
\end{figure}

Indeed we find $\mathbf E$ to be longitudinal and both,
$\mbox{curl}\, {\mathbf E}$ and ${\mathbf k}$, to be purely azimuthal
in the centre-plane between the sources.
Moreover the equality, Eq.~(\ref{dualamp}) is strikingly well obeyed by
our data, as visualised in Fig.~\ref{ampere}, which exemplifies the
situation for  $R = 8$.  This finding provides both support for
the superconductivity scenario and an {\it a posteriori} justification
of the underlying construction of the Abelian monopole current!

In the dual situation, the Cooper pairs from standard
superconductivity are replaced by a condensate of magnetic monopoles,
and their linear extension is now transcribed into a correlation
length, $\xi$. The latter can be retrieved either from the monopole
correlator\footnote{See DiGiacomo's lecture.} {\it or the coherence
length characteristic to the condensate wave function} as described by
the effective dual Ginzburg-Landau (GL) theory\cite{GL}. For
small
$\xi/\lambda$ ratios, we find ourselves in the deep dual London
limit, where $\mathbf k$ is directly connected to the chromo-electric
vector field, being related to the electric 
vector potential $\mathbf A$ through,
\beq
{\mathbf E} = \mbox{curl}\, \mathbf A.\label{vecpot}
\eeq  
We consider a colour electric string between charges $Q$
and $\overline{Q}$ placed on the $z$-axis, at $z = 0$ and $R$.  Let $x$
denote the radial distance of points from the $z$-axis.  Combining the
London equation in presence of this string,
\begin{equation}
\mbox{curl}\, {\mathbf k}  = - \frac{1}{\lambda^2} \left({\mathbf E} -
\frac{\Phi}{2\pi}\delta(x){\mathbf e_z}\right),
\label{dual}
\end{equation}
with the dual Amp\`ere law, Eq.~(\ref{dualamp}), 
one finds the relation for  the $\mathbf E$-component, longitudinal to
the vortex direction connecting the locations of $Q$ and $\overline{Q}$:
\begin{equation}
E_z(x)-\lambda^2\Delta_2 E_z(x) = \frac{\Phi}{2\pi} \delta^2(x), \label{besseleq}
\end{equation}
the solution of which is the profile function,
\begin{equation}
E_z(x)=\frac{\Phi}{2\pi\lambda^2} K_0 (x/\lambda).
\label{bessel}
\end{equation}
Here $K_0$ is the textbook Bessel function, and $\Phi$ denotes the electric
flux associated to the centre vortex.

\begin{figure}[ht]
\epsfxsize=12.5truecm
\centerline{\epsfbox{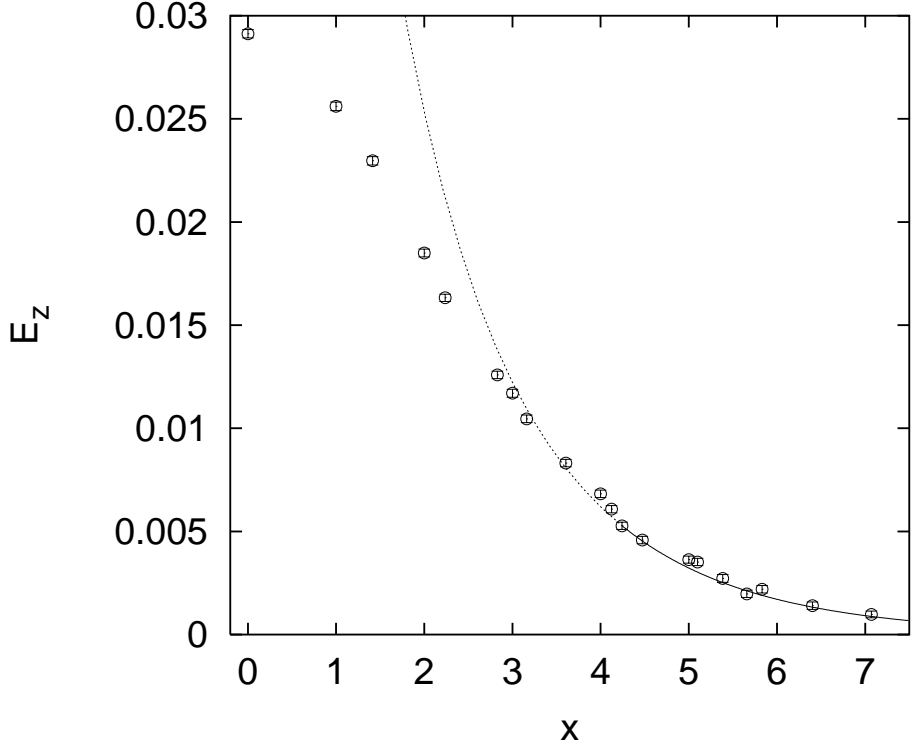}}
\caption{Fit of the  $E_z$-data to the London prediction, Eq.~(\protect\ref{bessel}).}
\label{deeplondon}
\end{figure}
\begin{table}[ht]
\caption{Fitting $E_z$ to Eq.~(\protect\ref{bessel}) at
  $R=8$, $T=6$ and various fit ranges. $\mu = (\lambda^2 k)^{-1/2}$
is the dual photon mass,  in units of the string tension $k$.}
\label{tab-fitbessel}
\begin{center}
\begin{tabular}{ccccc}\hline\hline
Fit range& $\Phi$ & $\lambda$  & $\mu$ & $\chi^2/{dof}$ \\ \hline
1\ldots   7.07 & 1.6(2) & 3.6(5)  & 1.54(21) & 1700 \\
2\ldots   7.07 & 1.2(1) & 2.4(1)  & 2.31(10) &  150 \\
3\ldots   7.07 & 1.2(1) & 2.0(1)  & 2.77(14) &    6 \\
3.1\ldots 7.07 & 1.3(1) & 2.0(1)  & 2.77(14) &    9 \\
3.6\ldots 7.07 & 1.4(1) & 1.9(1)  & 2.92(15) &    4 \\
4\ldots   7.07 & 1.5(1) & 1.88(6) & 2.95(09)  &    1.8 \\
4.2\ldots 7.07 &  1.4(1) &  1.82(7) & 3.05(12) & 1.2 \\
4.5\ldots 7.07 &  1.5(1) &  1.82(8) & 3.05(13)  & 1.1 \\
5\ldots   7.07 &  1.5(1) &  1.66(7) & 3.34(17) & 0.7 \\ \hline
\end{tabular}
\end{center}
\end{table}

For the $Q\overline{Q}$ separation $Ra = 8a\approx 0.7$~fm, which
is well within the flux tube
domain, we display the centre-plane distribution of $E_z$ in
Fig.~\ref{deeplondon}.  Our data reveals that these field distributions
are well described by the prediction of the dual London
theory, Eq.~(\ref{bessel}), iff $x$ is chosen sufficiently deep {\it in
vacuo}, see Table~\ref{tab-fitbessel}.
We emphasise that with appropriate lower bounds in the $x$-fit range we
gain stable values of $\Phi$ and $\lambda$. We are therefore in the
position to quote rather precise fit values from our dual London limit
analysis,
\begin{equation}
\Phi = 1.44(8), \quad\lambda  = 1.82(7). \label{londoneq}
\end{equation}
Note, however, that the value of the chromo-electric flux is far off
its (naively) anticipated size which is $\Phi = 1$ for our quarks,
carrying {\em one} unit of charge! This discrepancy is due to the
fact that the parametrisation overestimates the electrical field 
strength at small $x$.

\subsection{Ginzburg-Landau Analysis}
The failure of Eq.~(\ref{bessel}) to account for the entire set of
$E_z$-data hints at the existence of another mass scale, in addition
to the dual photon mass, $\lambda^{-1}$.  The extraction of this
second scale will now be tackled in the framework of the effective
dual GL-approach. 
In our specific axial geometry, the modulus of the wave function, $f$,
depends on $x$ alone, \beq \psi = \psi_\infty f(x){\rm e}^{i\theta(x)}.
\eeq Deep inside the vacuum, at $x = \infty$, $f$ is
normalised to {\it one}.  Apart from that, it is controlled by the
non-linear GL-equations\cite{tinkham},
\begin{eqnarray}
\label{GL_f}
f(x) &=& f(x)^3 + \xi^2\left[\left(\frac{1}{x}-\frac{2\pi
A_\theta(x)}{\Phi}\right)^2
-\frac{1}{x}\frac{d}{dx}\left(x\frac{d}{dx}\right)\right]f(x),\\
\label{GL_k} k_\theta(x) &=&
\frac{f(x)^2}{\lambda^2}\left[\frac{\Phi}{2\pi x} - A_\theta(x)\right].
\end{eqnarray}

Note that previous QCD analyses of these equations were based on
modelling the condensate wave function with  the ansatz,
\beq
\label{f-ansatz} f(x) = \tanh(x/\alpha)  \quad \mbox{with} \quad
\quad \alpha = \xi /\nu.   \label{small-x} 
\eeq

In the literature one finds $\nu$ treated as `fudge factor' with
{\it
ad hoc} value chosen to be {\it one}.  Mind that this arbitrariness in
$\nu$ affects the reliability of the value of $\xi$.
An examination reveals that the nonlinear character of
the GL-equations implies the  constraint, from the b.c. at $x = 0$:
\beq \nu =
 \sqrt{\frac{3}{8}}\left[1 -
 \frac{3\pi}{4\Phi}E_z(0)\alpha^2\right]^{-1/2}.  \eeq 

In order to acquire a reliable number for  $\nu$, one would evidently need
information about  $E_z(0)$,  safe from lattice artefacts ---
a difficult enterprise.

We will therefore opt for an alternative strategy by utilising the
 Monte Carlo data (on $\mathbf E$ and $\mathbf k$) to extract $f$
 directly from the GL-equations. To this end we choose {\it
 parametrisations} on $E_z$ and $k_\theta$ that respect the
 constraints from the {\it boundary conditions} on $f$.  In this way we gain
 access to the physical quantities of interest: $\xi$, $\lambda$, and
 $\Phi$, making full use of the structure of the effective
dual GL-equations in the continuum.

While for large $x$, $f$ should asymptotically reach the value
{\em one}, Eq.~(\ref{GL_f}) requires $f$ to vanish
linearly with $x$ at the centre of the vortex.

Let us remark that at this stage already, $\xi$ is likely to be
smaller than $\approx 4$: the reason being that the pure London ansatz
proved to be successful in the regime $x \geq 4.2$.  Thus our data
appear to provide no room for another length scale beyond this point.

To proceed from our data we must first determine \apot.  This is
achieved by simple integration over the chromo-electric flux: 
\beq
A_\theta(x) = \frac{1}{x}\int_0^x\! dx' x'E_z(x') = \frac{1}{2\pi x}\int\!
d^2\!f\, E_z ,\label{integration}\eeq
where $A_{\theta}$ complies with the limiting behaviours,
\beq 
A_\theta(x)
= \frac{\Phi}{2\pi x} \quad \quad (x\rightarrow \infty)
\quad \mbox{and}\quad A_\theta(x) = \frac{E_z(0)}{2}x \quad (x
\rightarrow 0).\eeq 

Remember that our analysis basically starts
 out from the second GL-equation, Eq.~(\ref{GL_k}) and determines the
 condensate wave function $f$ from fits to our data for $E_z$ (and
 induced $k_\theta $), by use of parametrisations satisfying the
 boundary conditions on $f$. Eq.~(\ref{GL_k}) can be solved for $f/\lambda$,
 \beq \frac{f(x)}{\lambda} = k^{1/2}_\theta(x) \left[\frac{\Phi}{2\pi x} -
A_\theta(x)\right]^{-1/2},\label{fdet} \eeq 
and cast by use of
Eqs.~(\ref{vecpot}) and (\ref{dual}) into the form
 \beq \label{constraint}
f^2(x)\int_x^{\infty}\! dx' x'E_z(x') = -\lambda^2 x \frac{dE_z}{dx}.
\eeq
\begin{figure}[t]
\epsfxsize=12.5truecm
\centerline{\epsfbox{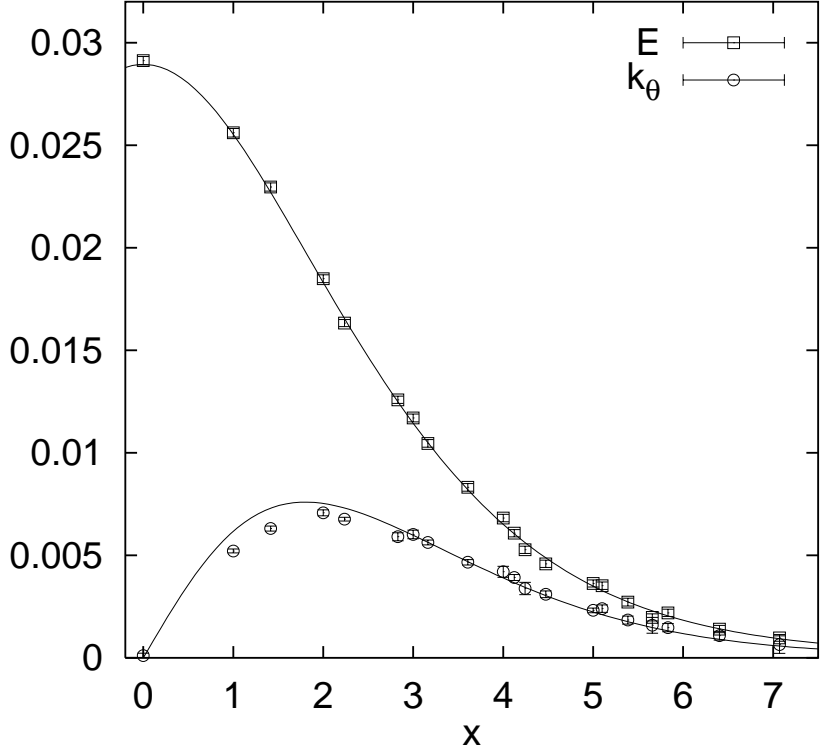}}
\caption{Fitting $E_z$ with the ansatz, Eq.~(\protect\ref{ansatz}),
 and comparing with 
data on $E_z$ and $k_\theta$.}\label{bcfite}
\end{figure}

We will now exploit this relation for the determination of $f(x)$.
Before we can reach this goal we must find a parametrisation for
$E_z(x)$ that meets the requirements induced through the boundary
conditions on $f$, as set by the GL-equations. Far away from the
vortex, where $f \rightarrow 1$, one recovers,
\begin{equation}
E_z(x)=- \lambda^2\frac{1}{x}\frac{d}{dx}x\frac{d E_z}{d x}.
\end{equation}
From the previous deep London limit solution, $K_0(x/\lambda)$,
one concludes the asymptotic behaviour $\sim \exp(-x/\lambda)$
for $x \rightarrow \infty$. A good parametrisation for $E_z$,
which is non singular throughout the entire $x$-interval
 reads therefore,
\begin{equation}
E_z(x)=\frac{\Phi}{4\pi C_a\lambda^2}\cosh^{-1}(x/\lambda) ,
\end{equation}
where the Catalan constant $C_a$ assures
the flux to be normalised to $\Phi$:
\begin{equation}
C_a=\sum_{j=0}^{\infty}\frac{\left(-1\right)^j}{\left(2j+1\right)^2}
=0.91596559\ldots.
\end{equation}
In the vicinity of $x=0$ we expect $f(x)= x/\alpha$.  This
induces, by Eq.~(\ref{constraint}), the differential equation
\begin{equation}
E_z(x) =\alpha^2\lambda^2\frac{1}{x}\frac{d}{d x}\frac{1}{x}
\frac{d E_z}{d x},
\quad \mbox{with solution} \quad
E_z(x)=\frac{\Phi}{\pi\delta^2}\exp\left(-\frac{x^2}{\delta^2}\right),
\end{equation}
where  $\delta=\sqrt{2\alpha\lambda}$. These considerations suggest 
the following global  four-parameter  ansatz  for our further fitting:
\begin{eqnarray}\label{ansatz}
E_z(x)&=&\frac{\Phi}{2\pi}\left[\frac{b}{2C_a\lambda^2}
\frac{1}{\cosh\left(x/\lambda\right)}
+\frac{2(1-b)}{\delta^2}\exp\left(-\frac{x^2}{\delta^2}\right)\right],
\label{epara}\\
k_{\theta}(x)&=&\frac{\Phi}{2\pi}\left[\frac{b}{2C_a\lambda^3}
\frac{\tanh(x/\lambda)}{\cosh(x/\lambda)}
+\frac{4(1-b)x}{\delta^4}\exp\left(-\frac{x^2}{\delta^2}\right)\right].
\end{eqnarray}
\begin{figure}[htb]
\epsfxsize=12.5truecm
\centerline{\epsfbox{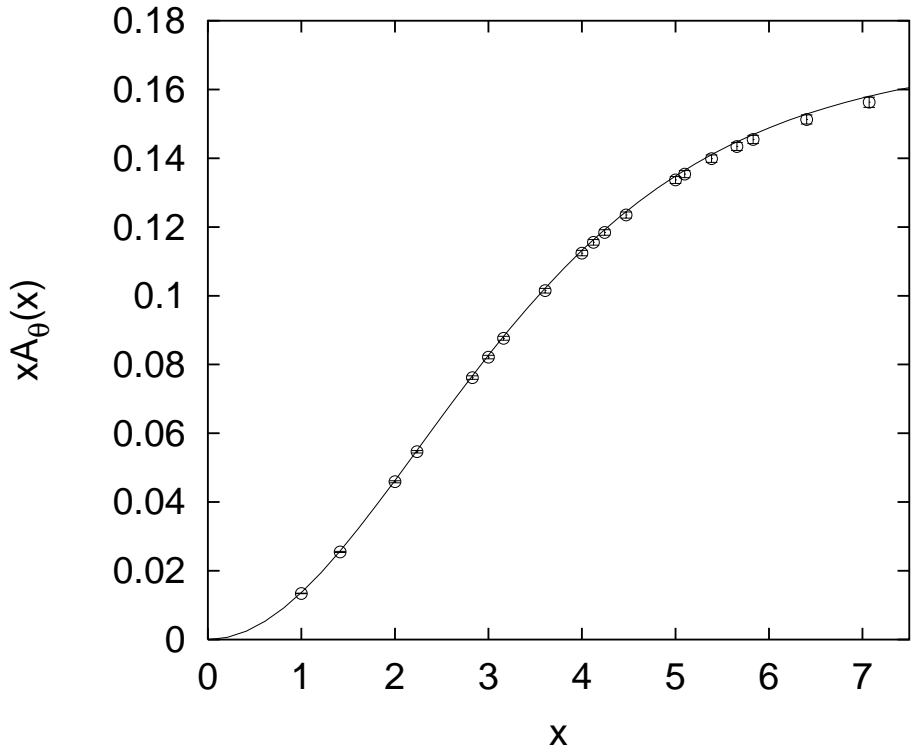}}
\caption{$xA_{\theta}$
from numerical integration and from the parametrisation, Eqs.~(\ref{ansatz}),
(\ref{values1}).}
\label{fig2}
\vskip 0.0truecm
\end{figure}

Fitting just the data for $E_z$ alone results in $\chi^2/N_{DF}=21.4/16$,
with the parameter values
\begin{equation}\label{values1}
\Phi=1.08(2),\quad \lambda=1.84(8),\quad \delta=3.28(23),
\quad b=0.71(6).
\end{equation}
Note that the value of $\Phi$ is now very close to the expectation,
yet $\lambda$ being fully consistent with the London limit result,
Eq.~(\ref{londoneq}). The quality of this fit on $E_z$ is exhibited in
Fig.~\ref{bcfite}.  However, this very figure also shows that the
$k_\theta$-data in the region $x \leq 2.2$ are not overly well
reproduced with the parameters quoted in Eq.~(\ref{values1}),
despite our convincing confirmation of
the dual Amp\'ere Law
on the lattice. This,
however, is not really unexpected since lattice artefacts are likely to
enter the game in this region.

\begin{figure}[htb]
\epsfxsize=12.5truecm
\centerline{\epsfbox{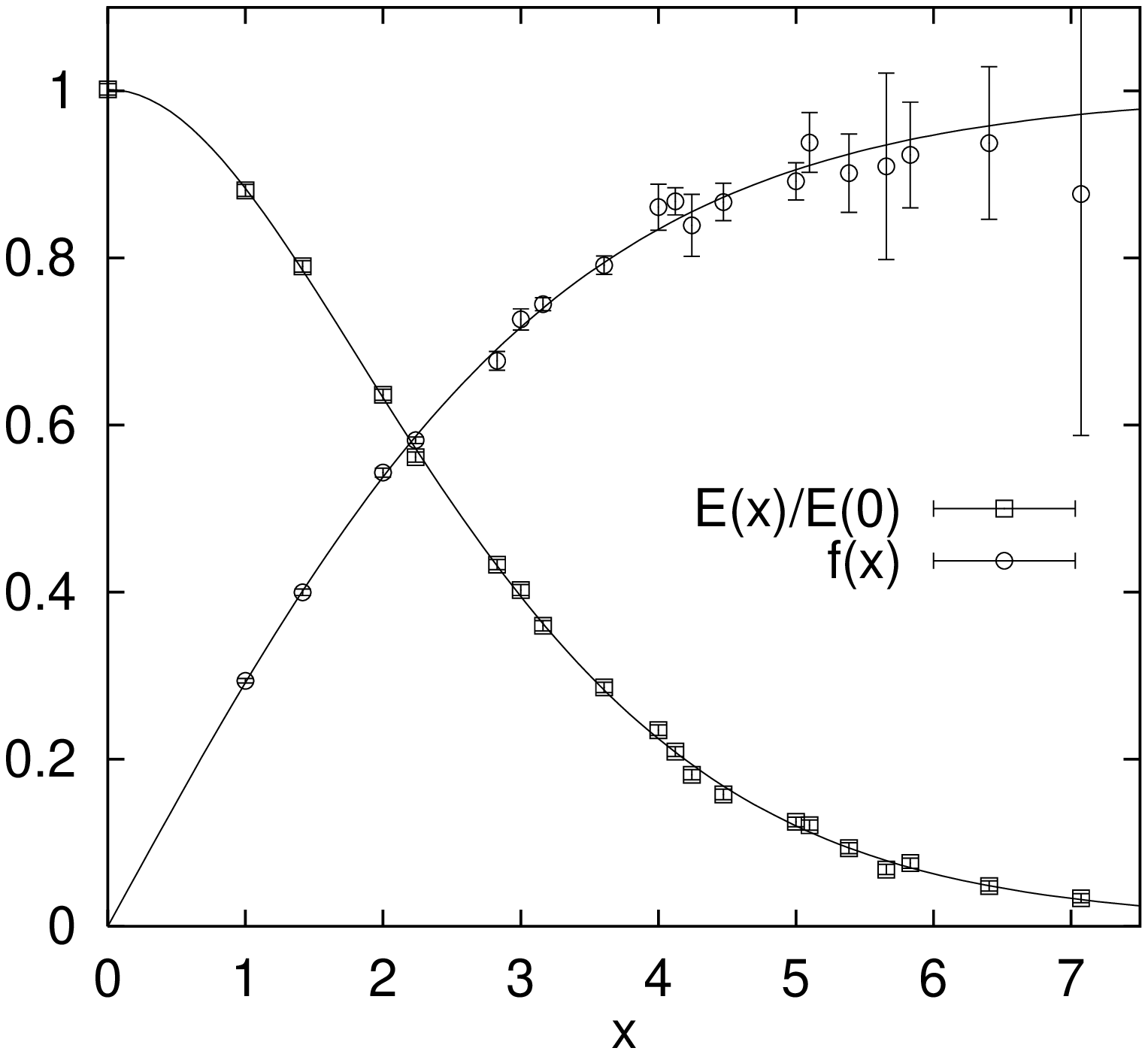}}
\caption{The boundary region as seen by $E_z$ and $f$.
$E_z$ as in Fig.~\protect\ref{bcfite} and $f$ from
Eq.~(\protect\ref{fdet}).}
\label{twoscales}
\end{figure}

To expose  the uncertainties from lattice artefacts, we allow next for
different values of $\delta$ (say $\delta_E$, $\delta_k$) in the $E_z$ and
$k_\theta$ parametrisations, Eq.~( \ref{ansatz}) and obtain a slightly
different set of parameter values,
\begin{equation} 
\Phi=1.10(2),\quad \lambda=1.99(5),\quad \delta_E=3.03(9),\quad
\delta_k=3.24(9),\quad b=0.67(3).
\end{equation}
from a combined  fit to the $E_z$ and $k_\theta$ distributions.
The curve in Fig.~\ref{ampere} corresponds to the above
parameter values. Although the fit has a reasonable value of
$\chi^2/N_{DF}=21.4/16$
it is physically not really sensible
since it goes along with   an unstable,  non-monotonic behaviour of
$f(x)$, as $\delta_E\neq\delta_k$. Nevertheless, it 
provides us with an
estimate of discretisation errors.

Before solving Eq.~(\ref{fdet}) for $f$, we numerically integrate
$E_z$, in order to obtain $xA_{\theta}$. A comparison between
numerically integrated data points and integrated
parametrisation, Eq.~(\ref{epara}), is displayed in Fig.~\ref{fig2}.
The result on $f$ is shown
in Fig.~\ref{twoscales}. The {\em rhs} of
Eq.~(\ref{fdet}) has been multiplied by $\lambda$ as obtained
from a two parameter fit of the data to a $\tanh (x/\alpha)$ ansatz. 
The parameter values are,
\begin{equation}\label{values2}
\alpha=3.33(5),\quad\lambda=1.62(2),\quad  (\chi^2/N_{DF}=10.0/17).
\end{equation}
Fig.~\ref{twoscales} visualises the two characteristic scales governing
the transverse flux tube profile, one being carried by $E_z$, the
other being borne by the wave function $f$.
Note that the statistical errors on $f$ explode in the region $x>4$,
which was indistinguishable from the London limit.

We decide to interprete the differences between the $\lambda$ values
above as a systematic uncertainty which we include into the
error of our estimate,
\beq
\lambda = 1.84^{+20}_{-24}.\label{lamres}
\eeq

\begin{figure}[htb]
\epsfxsize=12.5truecm
\centerline{\epsfbox{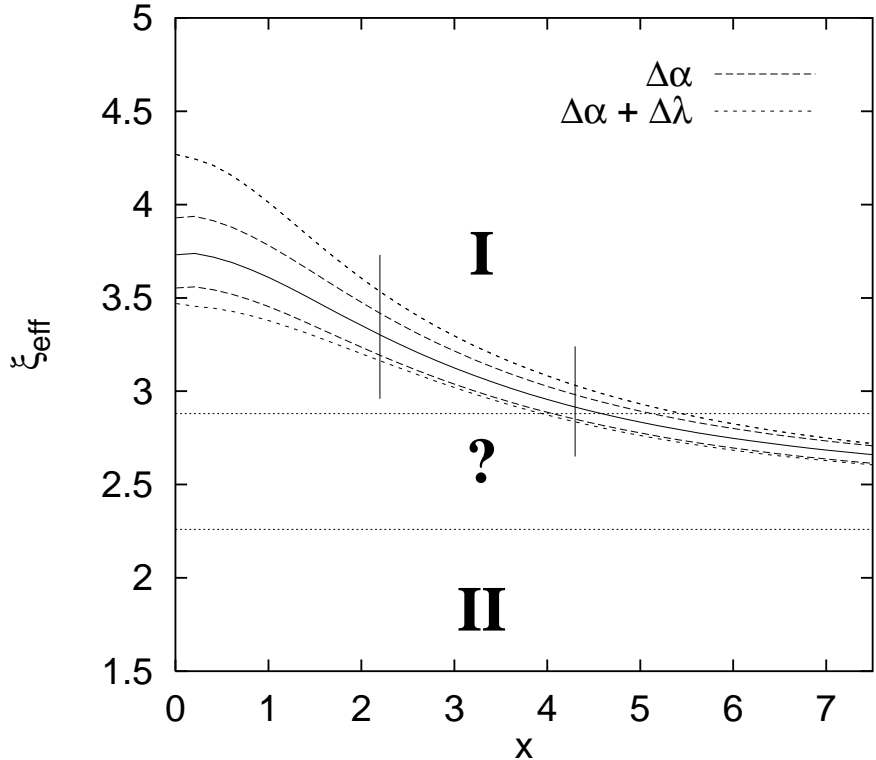}}
\caption{Effective $\xi$ as a function of $x$ with  error bands
      relating to  uncertainties in $\alpha$ and $\lambda$.}
\label{xieff}
\end{figure}
Let us finally compute $\xi$ from the first GL-equation,
Eq.~(\ref{GL_f}), by inserting the $\tanh (x/\alpha)$ ansatz for
$f(x)$, with $\alpha$ from the fit, Eq.~(\ref{values2}), and solving
 Eq.~(\ref{GL_f}) locally for $\xi_{\mbox{\scriptsize eff}}(x)$. It is of course not at all
guaranteed  that the numerical outcome of this procedure is
$x$-independent, as it should be: any deviation in $\xi_{\mbox{\scriptsize eff}}(x)$ from
a constant reflects  systematic uncertainties. In fact  we do
observe a 30 \% variation of $\xi$ as we travel through the
transition region, as depicted  in Fig.~\ref{xieff}.
The two error bands correspond to statistical and systematic
errors, respectively. This very figure
also indicates the window of observation, $2.2 < x < 4.2$, where we
expect to be sensitive to the structure of the wave function. 
Data obtained at smaller $x$ is unreliable due to lattice artefacts
while for large $x$, the uncertainty on $f$ and therefore $\xi$
certainly exceeds the error band suggested by the tanh-parametrisation.
Within this window, we find the effective $\xi$ to be 
\beq
\xi = 3.10^{+43}_{-35}. \label{xi}
\eeq
This estimate would indicate a GL-parameter $\kappa = \lambda/\xi =
0.59^{+13}_{-14}$
which is somewhat below the breakpoint $\kappa = 1/\sqrt{2}$ between type
II and type I superconductor as indicated
in Fig.~\ref{xieff}. Our estimate on $\sqrt{2}\lambda$ from
Eq.~(\ref{lamres}) is included as the ``?'' transition region.

\section{Discussion}
We have verified the validity of the dual Amp\`ere Law in maximal
Abelian gauge projection of SU(2) gauge theory.  In the London limit
analysis of the penetration region around an Abrikosov vortex built up
by a static $Q\overline{Q}$ pair, we have found a value of the dual photon
mass, $\mu a^{-1} = (\lambda a)^{-1}\approx 3\sqrt{\kappa}$
(Table~\ref{tab-fitbessel}),
which is well within the ball-park of the known
SU(2) glueball masses. In a
Ginzburg-Landau analysis, the value of the chromo-electric flux $\Phi $
is in qualitative agreement with the expected value {\it one} while the
$\lambda$-value is in accord with the penetration length from the
London limit analysis.  

We find a small window, $2.2<x <4.2$, within the transition region between
vortex and superconducting vacuum from which we can view the GL-wave
function and determine the coherence length of the chromo-magnetic
condensate. While the wave function varies by about a factor {\em two}
within this window, the
effective value of $\xi$  is constant within 10 \%,
yielding the value quoted in Eq.~(\ref{xi}).  This appears to be
distinctly above the limit $\sqrt{2}\lambda$, indicating that we are
faced with a type I superconductor scenario, contrary to previous
expectations.  

This can only be taken as a tentative conclusion. In order to
settle the issue one might study flux profiles with more than
{\it one} unit of chromo-electric flux involved.  It goes without
saying that a final answer would require a scaling study as well.

\section*{Acknowledgements}
K.~S.\ thanks Prof.\ T.~Suzuki and his team for their splendid
organisation and the inspiring atmosphere of the YKIS97 seminar at the
Yukawa Institute at Kyoto University.  We acknowledge support by the
DFG (grants Schi 257/1-4, Schi 257/3-2, Ba 1564/3-1 and Ba 1564/3-2)
and the Wuppertal DFG-Graduiertenkolleg ``Feldtheoretische und
numerische Methoden in der Statistischen und Elementarteilchenphysik''.
The computations were mostly carried out on the Wuppertal CM-5 system,
whose disk array was funded by DFG.

\end{document}